\documentclass[12pt]{article}
\usepackage{epsfig}

\def\Title#1{\begin{center} {\Large {\bf #1} } \end{center}}

\begin{document}

\Title{FROM ACCELERATORS TO ASTEROIDS*}

\bigskip\bigskip


\begin{raggedright}  

{\it Martin L. Perl \\
Stanford Linear Accelerator Center \\
Stanford University \\
Stanford, California 94309 }
\bigskip\bigskip
\end{raggedright}

\vspace{4 in}

* Written version of talk presented at the Cosmic Genesis and Fundamental Physics Workshop, Sonoma State University, October 28-30, 1999

\pagebreak

\section{FROM ACCELERATORS TO ASTEROIDS:\\EXTENDING THE REACH OF PARTICLE\\ PHYSICS}

This Workshop on Cosmic Genesis and Fundamental Physics includes many experimental and theoretical subjects in astrophysics, astronomy, and cosmology, Fig. 1. 

\begin{figure}[htb]
\begin{center}
\epsfig{file=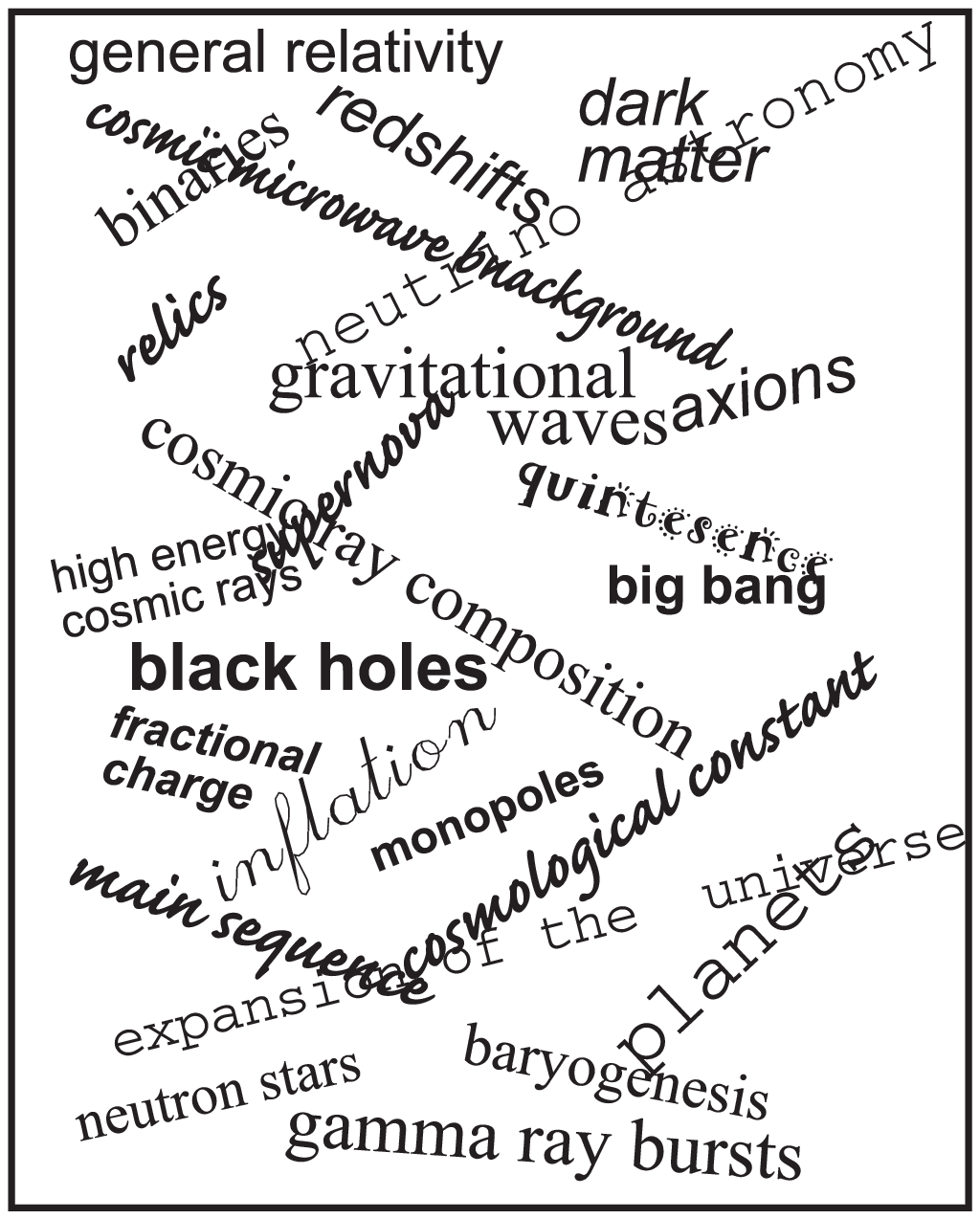,width=11cm}
\caption{Some subjects in astrophysics, astronomy and cosmology.}
\end{center}
\end{figure}

In this talk I take up one subject: {\it how can the experimental reach of traditional elementary particle physics be extended by using the methods and findings of experimental and observational astrophysics and astronomy?} We need the broadest obtainable reach because particle physics does not have known experimental boundaries.

\begin{itemize}
\item There is no known theoretical upper limit on the masses of the particles that we might seek.   
\item There are few theoretical limitations on new phenomena or new forces that might appear at energies not yet reached.   
\item There is no smallest non-zero mass that we consider uninteresting.   
\item There is no smallest distance that we consider uninteresting.   
\item There is no large distance that we consider uninteresting for the testing of our understanding of  some of  the elementary forces. 
\end{itemize}

In this talk I discuss the reach of traditional methods of elementary particle physics in five broad experimental areas. For each of these areas I inquire how that reach  is, or might be, extended by experimental and observational astrophysics and astronomy. The five areas are:

\begin{itemize}
\item Searches for particles with very large masses and measuring those masses.   
\item Searches for particles with very small masses and measuring those masses.   
\item Searches for new types of particles.   
\item Searches for unexpected behavior of the known forces or for new forces.   
\item Searches for new phenomena at very high energies.
\end{itemize}

\section{SEARCHES FOR PARTICLES WITH VERY LARGE MASSES }

\subsection{Dreams and goals}

The state of knowledge in any science depends upon the state of the experimental technology. I illustrate this in Fig. 2 for particle physics by using  a mass scale extending from $10^{-6}$ to $10^{27}$ eV/c$^{2}$, 33 decades. There is no significance to the upper limit; if I was willing to decrease the readability of the figure, I could have added, say, ten more decades to allow particles with masses in the kilogram range. Why not?

\begin{figure}[htb]
\begin{center}
\epsfig{file=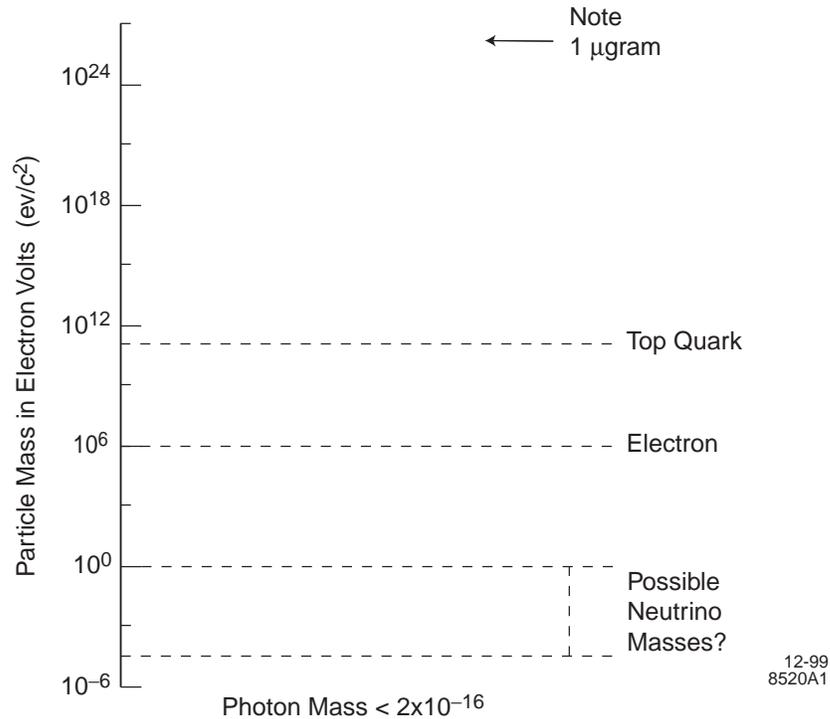,width=11cm}
\caption{The range of the masses of the known elementary particles.}
\end{center}
\end{figure}

Except for the photon with a mass less than $2\times10^{-16}$ eV/c$^{2}$, perhaps zero, and the gluon whose mass may be zero, the masses of the known elementary particles are included in the lower 17 decades of this figure. The finding of particles with masses extending over 17 decades is a remarkable achievement of the experimenters. Still, we recognize the limitation of the existing technology of atomic, nuclear, and particle physics. One of our experimental dreams is to search for more  massive particles; they must exist. I do not believe that we have been so smart and so lucky to have discovered  all existing particles in the twentieth century.

\subsection{Accelerator searches with known technology}

How much higher can we probe with the accelerators now under construction or at least being considered? This is answered in Fig. 3. With e$^{+}$e$^{-}$ colliders and  $\mu$$^{+}$$\mu$$^{-}$  colliders we can directly probe to about $10^{12}$ eV/c$^{2}$, a TeV/c$^{2}$. The Large Hadron Collider now being constructed may directly probe to several TeV/c$^{2}$. And a 100 TeV/c$^{2}$ on 100 TeV/c$^{2}$ Very Large Hadron Collider would directly probe to masses of several tens of TeV/c$^{2}$. 

\begin{figure}[htb]
\begin{center}
\epsfig{file=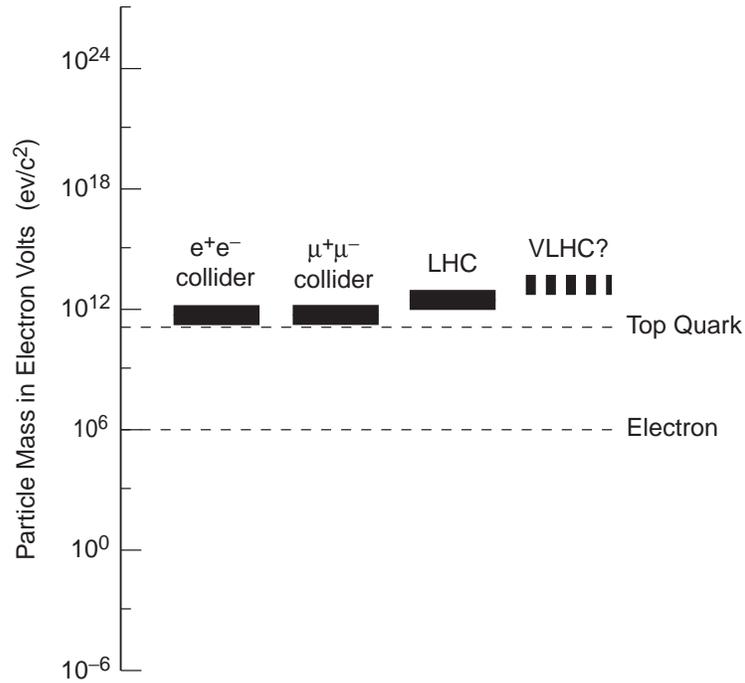,width=11cm}
\caption{Upper limits of direct mass searches for accelerators being constructed or proposed.}
\end{center}
\end{figure}

Of course we are smarter than that. We can indirectly probe to higher masses by looking for the effects of the low energy tail of s-channel resonances. As sketched in Fig. 4 this might extend the searches to $10^{14}$ eV/c$^{2}$. Then we will have extended the upper limit on mass searches by about $10^{3}$. With respect to what we now know this will be a great technical accomplishment. But with respect to the mass scale on these figures and with respect to our dreams, we have much further to go.

\begin{figure}[htb]
\begin{center}
\epsfig{file=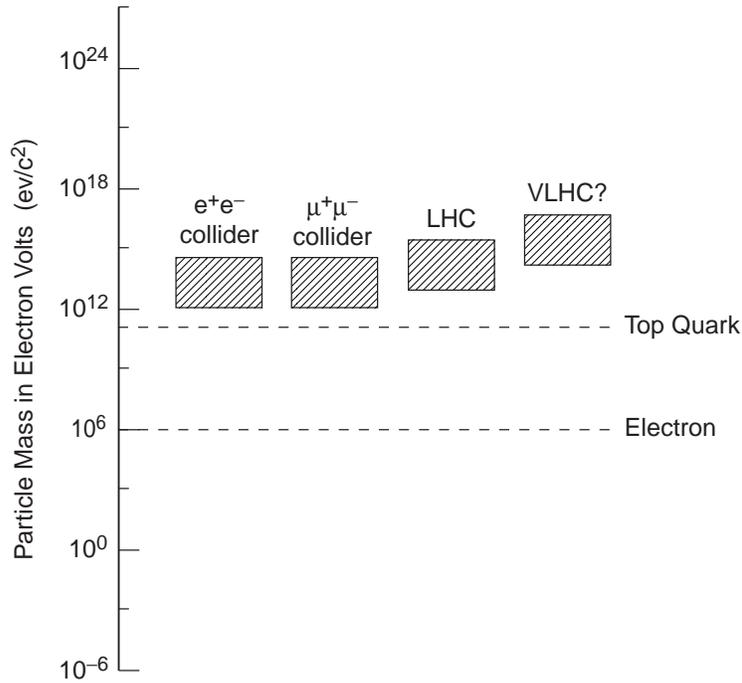,width=11cm}
\caption{Upper limits of indirect mass searches for accelerators being constructed or proposed.}
\end{center}
\end{figure}

\subsection{Accelerator searches with future technology}

I am an optimist; the upper limits in Fig. 4 will not be the end of what we will do with accelerators. One or two hundred years from now, the present technology of electron and muon and hadron colliders will seem primitive - as primitive as the Cockcroft-Walton accelerator seems to us. But no one knows what that accelerator future technology will be, nor what energies will be achieved. 

\subsection{Mass spectrometer searches for unusually heavy atoms} 

There is another traditional method for searching for massive elementary particles, a non-accelerator method. Mass spectrometry is used to look for unusually heavy atoms \cite{heavyatom}. For example, suppose there is a massive, positively charged particle called X$^{+}$. Then X$^{+}$e$^{-}$, a massive analog  to the hydrogen atom, would exist. More generally, a massive, negatively charged particle, X$^{-}$, could be incorporated in an atom, for example He$^{++}$ X$^{-}$e$^{-}$. And of course if the X partakes of the strong interaction then massive isotopes of some nuclei will exist. The present experimental upper limit on such searches is shown in Fig. 5 \cite{heavyatom}.

\begin{figure}[htb]
\begin{center}
\epsfig{file=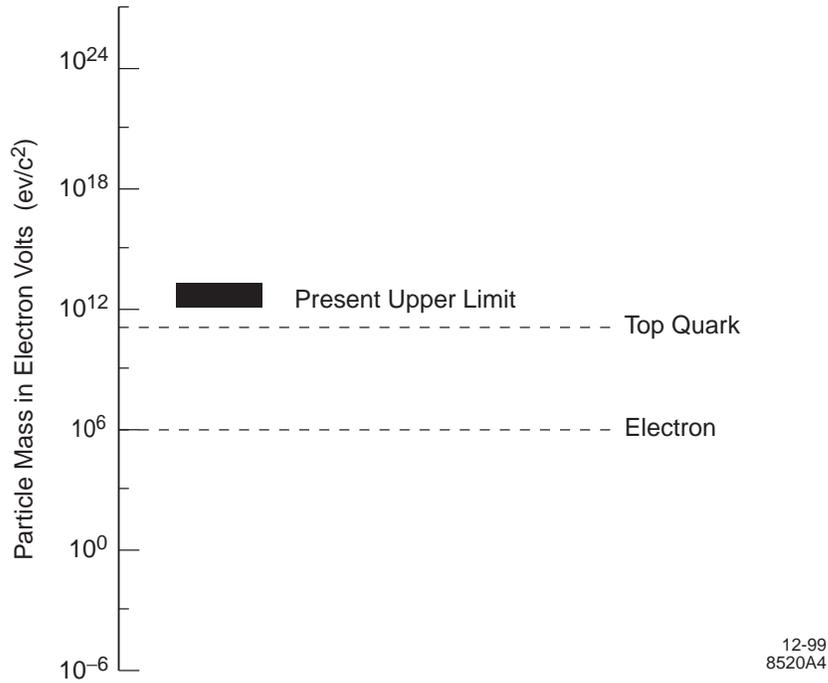,width=11cm}
\caption{Upper limit of searches for massive particles using mass spectrometry to look for unusually heavy atoms.}
\end{center}
\end{figure}

\subsection{Non-traditional concepts for very massive particle \\ searches}

How can we look for particles with masses above the upper limits in Figs. 4 and 5? Only by going to non-traditional concepts for searching for massive particles. I know of two such concepts, perhaps there are more.

First, there is the very general concept that  the study of extremely high energy cosmic rays - charged particles, photons, and neutrinos - might give clues to the existence of a very massive particle. Thus, there have been speculations that the charged particles with energies above $10^{19}$ eV/c$^{2}$ might come from the decay of a very massive particle, particularly one whose properties enable it to travel through space more easily than protons \cite{spaceparticle}. Here then is a crucial area where astrophysical observations could make a seminal contribution to elementary particle physics.

The second, non-traditional concept for very massive particle searches is much more limited in mass range. As described next, we have proposed a liquid drop search method \cite{per} for particles in the $10^{13}$ to $10^{17}$ GeV/c$^{2}$ mass range.

\section{SEARCHES FOR MASSIVE PARTICLES \\ PRODUCED IN THE EARLY UNIVERSE}

\subsection{Search motivation}

One can search in bulk matter for a class of very massive particles using a falling drop method \cite{per}. The criteria for particles in this class are:

\begin{enumerate}
\item Mass in the range of $10^{13}$ to $10^{17}$ GeV/c$^{2}$.
\item These particles would have to been produced in the early universe and be present  in the solar system.
\item Stable.
\item Charged or bound by the strong interaction to a stable charged particle.
\end{enumerate}
\subsection{Liquid drop search method for massive particles}

This method depends upon some mass relationships.  The mass of a 6 $\mu$m diameter drop with a typical mineral suspension of density 1.4 grams/cm$^{3}$ is

\vspace{.15 in}
 
$m_{drop}= 1.6\times10^{-10}$ grams.

\vspace{.15 in}
 
Since 

\vspace{.15 in}
 
1 GeV/c$^{2} = 1.8\times10^{-24}$ grams ,

\vspace{.15 in}
 
$ m_{drop}$= 10$^{14}$ GeV/c$^{2}$.

\vspace{.15 in}
 
Thus our smaller drops have a mass equal to or less than particles that might exist in the interesting mass range of 10$^{14}$ GeV/c$^{2}$ and above.

\begin{figure}[htb]
\begin{center}
\epsfig{file=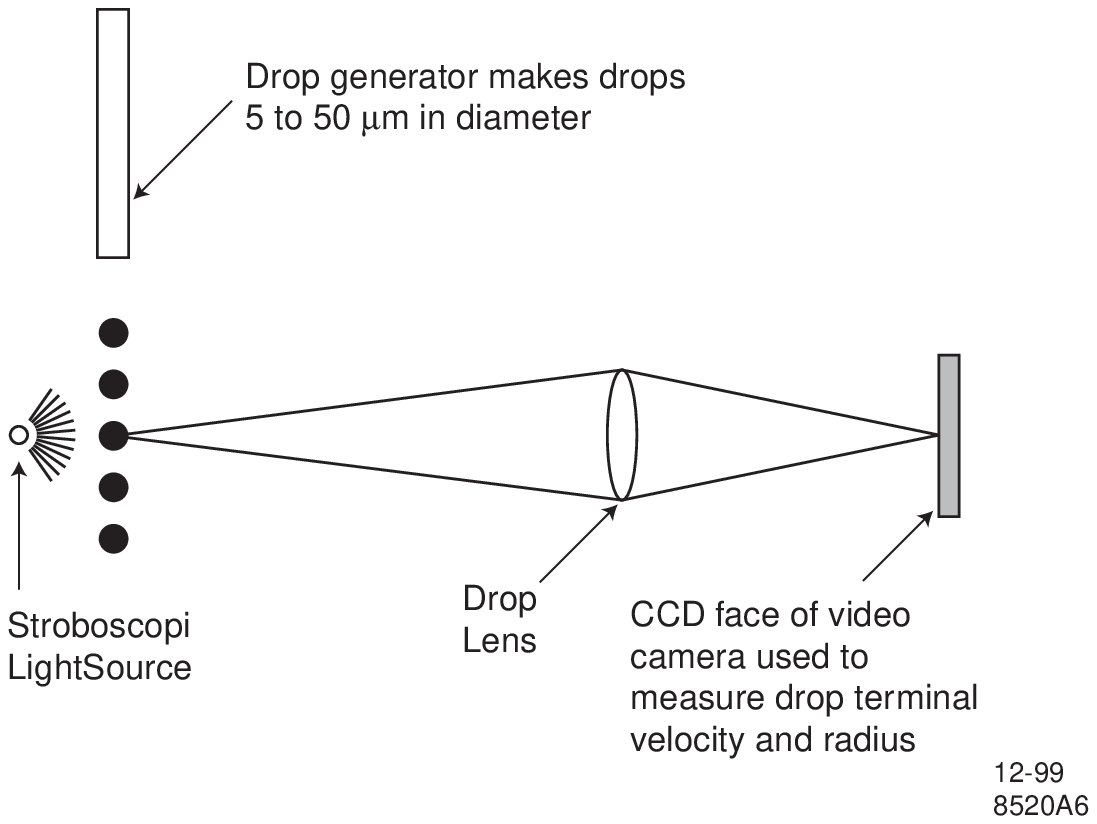,width=11cm}
\caption{Schematic illustration of liquid drop apparatus for searching for massive particles.}
\end{center}
\end{figure}

Consider an apparatus that measures the terminal velocity of drops falling in air, Fig. 6. A drop of mass m has terminal velocity $v(m)$:

\vspace{.15 in}
 
$v(m) = mg/6\pi\eta r$.

\vspace{.15 in}
 
Where $g$ is the acceleration of gravity, $\eta$ is the viscosity of air, and $r$ is the drop radius.  Suppose a drop also contains an elementary particle of mass M, then the terminal velocity is 

\vspace{.15 in}
 
$v(m+M) = (m+M) g/6\pi\eta r$.

\vspace{.15 in}
 
Figure 7, an illustrative plot of number of drops $dN/dv$ versus $v$, shows what we hope to see:  a very large peak at $v(m)$ and a relatively very small peak at $v(m+M)$.  Our ability to detect the $v(m+M)$ peak depends on the abundance of the massive particle and on the width and tails of the $v(m)$ peak.  As a first estimate we believe we can separate the $v(m)$ and $v(m+M)$ peaks if $M> m$.

\begin{figure}[htb]
\begin{center}
\epsfig{file=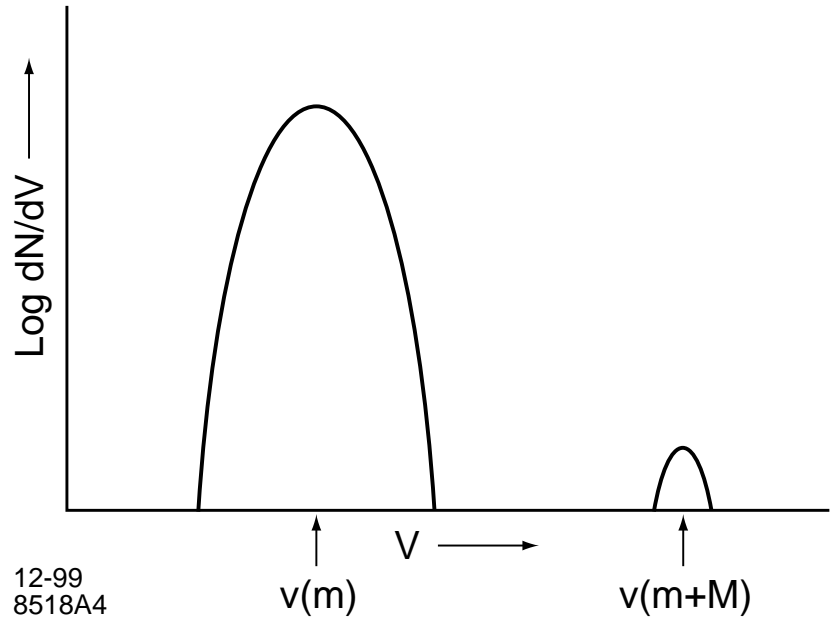,width=11cm}
\caption{Illustrative plot of number of drops $dN/dv$ versus $v$, with a very large peak at $v(m)$ and a relatively very small peak at $v(m+M)$.}
\end{center}
\end{figure}

\subsection{Lower and upper mass limits}

The lower mass limit of the search method is determined by the rough requirement $M\geq m$ and by the minimum size drops we can use in a practical experiment. We can probably reliably produce drops with 4 $\mu$m diameter, giving a lower limit on $M$ of about 10$^{13}$ GeV/c$^{2}$.  We do not see how to extend this method to yet smaller drops:  it may be difficult to make such drops reliably, it will be difficult to get reliable measurements of the drop radius r, and it will be difficult to search through large amounts of material.  

As discussed in  Ref. \cite{per}, the mass limit comes from the necessity in this search method of the massive particle remaining bound in ordinary matter while in Earth's gravitational field.  There must exist a binding force $F_{b}$ between the particle and the drop's ordinary matter so that $F_{b}$ is larger than the gravitational force on the particle, $Mg$.  The straightforward binding mechanism is electric charge.  We suppose the particle is charged or is bound by the strong force to a charged particle.  To estimate $F_{b}$, we suppose (a) the massive particle has an electric charge $e$, where $e$ is the electron charge, (b) the binding energy to the ordinary matter is about 1 eV, and (c) $F_{b}$ extends over about 10$^{-10}$m. Then $F_{b}$ is about 1.6$\times10^{-9}$ nt, and $M$ must be less than $F_{b}/g=1.6\times10^{-10} kg = 10^{17}$ GeV/c$^{2}$.

Hence this proposed search for massive stable particles with electric charge could extend from 10$^{13}$ to 10$^{17}$ GeV/c$^{2}$.  There is certainly some optimism in the calculation of these limits.  The lower limit might not be quite so low if it proves to be difficult to use drops of less than 6 $\mu$m diameter.  The upper limit might not be quite so high if the particle has fractional electric charge or we have been too generous in estimating the strength of $F_{b}$.

\subsection{ Near-term goals for searches for very massive particles}

In the course of developing this search method we will use a terrestrial mineral sample.  But the geological history of the earth is complicated and particles in the 10$^{13}$ to 10$^{17}$ GeV/c$^{2}$ mass ranges may have long since moved to the earth's center.

The best materials for very massive particle searches are meteorites from asteroids and it is here that we shall put our first serious effort.  Unfortunately, there is a problem with the upper mass limit when searching meteorites.  As pointed out by Jean and Longo \cite{jeo}, when meteorites enter the atmosphere they slow down, the deceleration force may be 100g to 1000g.  Therefore the more massive particles will not stay in the meteorite.

\subsection{Twenty-five year goals for searches for very massive \\  particles}

The organizers of this Workshop asked the participants to set twenty-five goals for their research interests. Our goal is to overcome the meteorite deceleration problem; there are two solutions.  One solution is to bring back asteroid samples by a small acceleration and small deceleration orbit, perhaps keeping the acceleration or deceleration to less than 10g.

The other solution, grand and exciting, is to send the massive particle search apparatus to an asteroid, carrying out the search on the asteroid.  There are three great advantages to this solution.
\begin{itemize}
\item There are no particle loss problems from acceleration or deceleration.
\item Since $g_{asteroid}<<g_{earth}$ the upper mass limit for searches is increased.
\item Since $g_{asteroid}$ is relatively small, very massive particles may lie on the surface.
\end{itemize}
We don't know if we can turn this dream into a reality, the technical problems are hard, but we don't know of any other way to search for very massive particles.

\section{SEARCHING FOR STABLE NEUTRAL\\PARTICLES AND MEASURING THEIR\\ MASSES}

There are two general methods for measuring the masses of elementary particles.

\vspace{.15 in}
 
{\bf Decay Method:  }If the particle decays, the mass can often be obtained by measuring the four-momenta of the decay products. This can be done even if one of the particles is a neutrino and is not detected.

\vspace{.15 in}
 
{\bf Four-momentum Method: } If the particle is stable or metastable and charged, then its mass can be determined by measurement of its orbit in a magnetic field and by measurement of its energy.
 
\vspace{.15 in}

These mass determination methods are doubly important because they are also used extensively for searching for new particles.

Unfortunately, {\it there is no general method for measuring the masses of stable neutral particles}. There are some special methods such as those used in determining the neutron mass. Also, if the neutral particle is one of the decay products of a known particle, it may be possible to find it and to measure its mass. However, as shown in Fig. 8, we have scanty knowledge of the masses of the {\it known} light, neutral particles.

\begin{figure}[htb]
\begin{center}
\epsfig{file=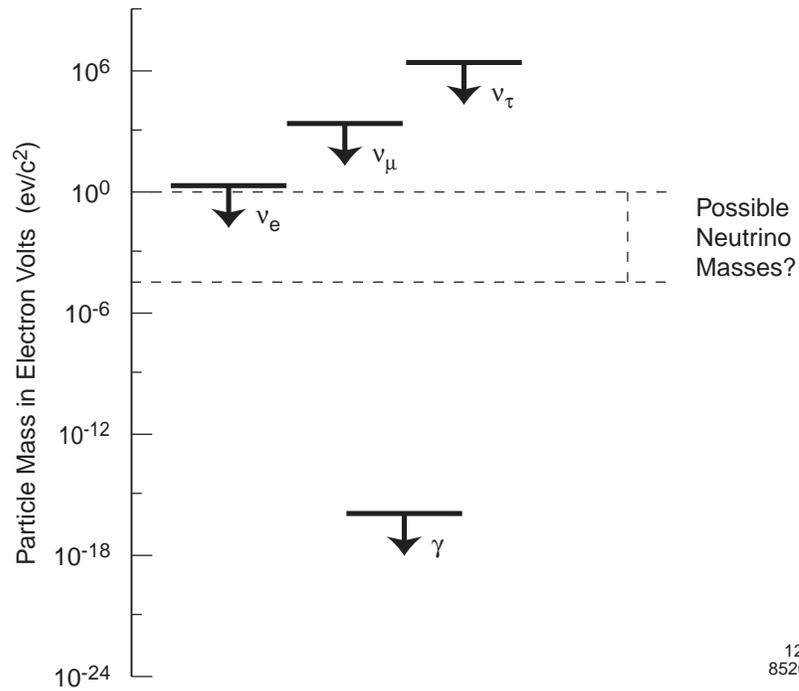,width=11cm}
\caption{Our scanty knowledge of the masses of the known, light, neutral particles.}
\end{center}
\end{figure}

{\it It is even more unfortunate that there is no general method for searching for unknown, stable, neutral particles}. There are some special methods such as those used in axion searches, Sec. 5.3. And if the neutral particle partakes of the weak interaction, we would have detected it in the decays of K mesons or $\tau$ leptons or Z$^{0}$'s. But if it is  a peculiar particle such as the axion or the graviton, it would not have been detected. Therefore we continue to rely on special methods for finding unknown, stable, neutral particles. It is to be hoped that the study of astrophysical phenomena will provide such methods; indeed the comparison of the behavior of stars with stellar theory has already provided valuable limits on what neutral particles can exist.

\section{SEARCHES FOR NEW TYPES OF PARTICLES}

Astrophysical and astronomical experiments and observations offer the most promise for the discovery of six new kinds of particles:

\begin{itemize}
\item Very massive particles, a topic already discussed.
\item Supersymmetric partners of known particles.
\item Dark matter.
\item Axions.
\item Magnetic monopoles.
\item Particles with fractional electric charge.
\end{itemize}

\subsection{ Searches for supersymmetric partners of known \\ particles.}

There have been so many calculations on the expected properties and interactions of supersymmetric partners of the known particles, there have been so many searches for these particles, there have been so many papers: but we have no confirmed evidence for their existence and we have no definitive experimental evidence for the reality of supersymmetric theory \cite{super}. It is surprising to me that so many hopes and dreams are still attached to supersymmetric theory.  There are a few reasons for this persistence. First, there is no convincing substitute theory. Second, if the theory were applicable to the real world it would explain numerous observations in particle physics. And third, there is always the hope that the supersymmetric partners will be discovered at higher energies. 

I can argue on either side of this hope for vindication of supersymmetric theory at higher energies. The pessimistic argument is that the energies reached by our present particle physics technology have been sufficient to find many members of the three classes of elementary particles, the leptons, the quarks, and the force-carrying particles; why should higher energy be required to find at least one supersymmetric partner? The optimistic argument is based on the history of that technology: higher energy has led to discovery.

Taking up the optimistic argument, I believe that conventional accelerator search methods are best suited for the discovery of supersymmetric particles. It may be that clues to the existence of a supersymmetric partner may be found in  astrophysical or astronomical observations, but it seems to me that it will be difficult to confirm the existence and to extract the properties.

In the end I am skeptical about the reality of a supersymmetric particle physics world. The searches and the calculations should certainly continues but we should not allow supersymmetric theory to blind us to the possibility of other ideas, we should not let supersymmetric theory prevent our taking fresh looks at accelerator and non-accelerator data.

\subsection{Searches for dark matter}

Experimental and observational research on dark matter involves two vast and intertwined questions:

\begin{enumerate}
\item{\bf How much dark matter exists and what is its distribution? There has been a great deal of progress using a number of ingenious astrophysical and astronomical methods \cite{darkmatter}. These methods include:}
\begin{itemize}
\item Studies of the internal dynamics of galaxies.
\item Studies of the dynamics of larger structures.
\item Dark matter tomography.
\end{itemize}

Of course the conventional techniques of particle physics have no contributions to make in answering this dark matter quantity and distribution question.

\item{\bf What is dark matter? There have been many proposals for the nature of dark matter: neutrinos, axions, weakly interacting massive particles (WIMPS), unknown new particles, non-luminous baryonic matter such as dwarf stars. There are two methods being used to try to elucidate the nature of dark matter.}
\begin{itemize}
\item One method depends upon the detection of the collision of terrestrial nucleons with an assumed gas of dark matter particles as the earth moves through that particle gas. Usually the dynamics of the collision are calculated  assuming the dark matter consists of massive, neutral, weakly-interacting, supersymmetric particles. This method, in a way, is an extension of traditional particle physics technology; however it illustrates how astrophysical questions have broadened that technology.
\item The other method depends upon annihilation of a dark matter particle with a dark matter antiparticle, and detection of that annihilation by a space-observing particle detector. The simplest example is an annihilation channel consisting of two gamma rays; the gamma rays would then be monoenergetic and thus detected. In the planning for such a search it is usually assumed that the dark matter would be concentrated at the center of a massive body such as the sun; this assumption increases the annihilation rate.
\end{itemize}

However we desperately need new experimental and observational methods to unravel the dark matter mystery. Is it possible that such a new method can come from the traditional technologies of atomic or nuclear or particle physics?

\item{\bf There is a third question that I must add before moving onto the next subject. Is it possible that our basic ideas about dark matter are wrong? That is, the observations are correct but the interpretation as matter is wrong. Might the dark matter concept be the phlogiston of the seventeenth century or the ether of the nineteenth century?}
\end{enumerate}

\subsection{Searches for axions}

The difficulty in searching for axions - small mass, neutral, weakly-interacting particles-illustrates the discussion in Sec. 4. We do not have a systematic way to search for small mass, neutral, weakly-interacting particles. Many ingenious techniques have been used to search for axions, but so far the axion is elusive \cite{axion}. Figure 9 shows the unexplored regions in $g_{A\gamma}$--$m_A$ space where $g_A$$\gamma$ is the axion coupling constant and $m_A$ is the axion mass. 

\begin{figure}[htb]
\begin{center}
\epsfig{file=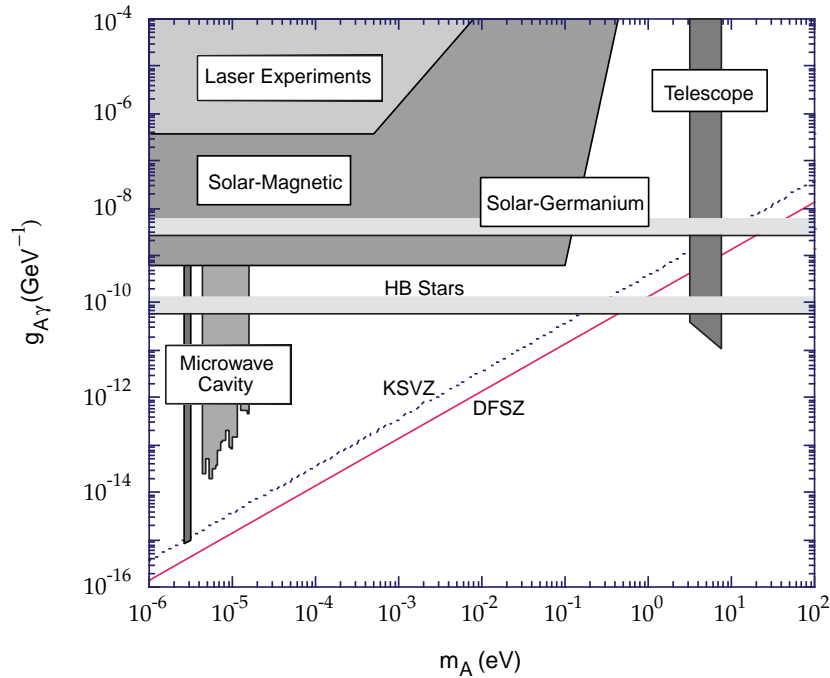,width=11cm}
\caption{Unexplored axion search regions in $g_{A\gamma}$--$m_A$ space where $g_{A\gamma}$ is the axion coupling constant and $m_A$ is the axion mass \cite{axion}.}
\end{center}
\end{figure}

\subsection{Searches for magnetic monopoles}

There are two different general methods of searching for magnetic monopoles. In one method the experimenter searches for a flux of monopoles entering the earth's atmosphere. A variety of techniques is  used: magnetic induction, ionization detection, Cerenkov radiation detection \cite{monopole}. {\it There are no confirmed discoveries of monopoles.} Figure 10 shows the measured upper limits on monopole fluxes compiled by B. C. Choudhary \cite{choudhary}.

The other general method is searching for monopoles in bulk matter; the material is passed through a superconducting coil with a squid for detection. There are two recent extensive searches, {\it both with null results.}; Jeon and Longo \cite{jeo} searched through 331 kg of material including 112 kg of meteorites,  Kovalik and Kirschvinki searched through 643 kg of rock and 180 kg of sea water \cite{kov}. The 90 percent confidence upper limit on the existence of monopoles in these materials is about {\it 10$^{-29}$ monopoles/nucleon}, probably the smallest upper limit on the abundance of hypothetical particles.

\begin{figure}[htb]
\begin{center}
\epsfig{file=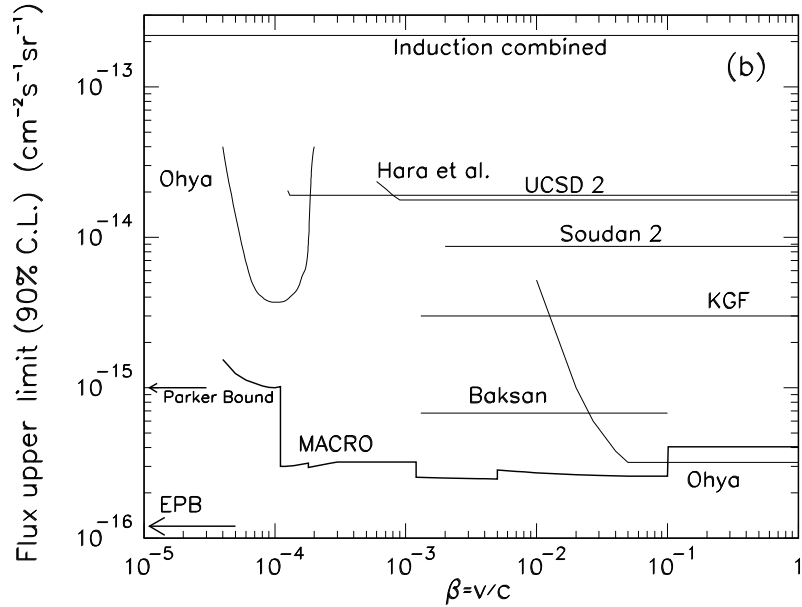,width=11cm}
\caption{Measured upper limits on monopole fluxes compiled \cite{choudhary}.}
\end{center}
\end{figure}

\subsection{Particles with fractional electric charge}

There are three ways of searching  for free elementary particles with fractional electric charge. By free I mean particles that can be isolated, such as leptons and photons, in contrast to quarks that conventional theory holds to have fractional charge but also holds to be bound into hadrons. Of course it is possible that very rarely a single quark exists in isolation, and such quarks are included in any search for free, fractional charge particles. The three search methods are:

\subsubsection{ The three search methods}

\begin{itemize}

\item {\bf Searches using accelerators:  } Accelerator searches for fractional charge particles are straightforward and many have been carried out \cite{lyo,jone,gur}. But they are limited in the mass range of the search by the maximum accelerator energy. Another problem is that the production cross section is unknown, so that a null search result does not rule out the existence of a fractional charge particle in that mass range \cite{perllee}. {\it There is no confirmed evidence from accelerator searches for the existence of fractional charge particles.}  Still it is certainly worthwhile to search for fractional charge particles when higher energy accelerators are put into operation.

\item {\bf Searches in cosmic rays:  } There are two different concepts behind cosmic ray searches \cite{lyo,jone,perllee}. In one concept it is assumed that the interactions of a primary cosmic ray in the atmosphere makes a fractional charge particle. This is subject to the same uncertainties  as accelerator searches. Higher energy compared to current accelerators is available but then the flux is very small. The other concept assumes that fractional charge particles are produced somewhere in space and impinge upon the earth. {\it There is no confirmed evidence in cosmic ray searches for the existence of fractional charge particles.} A recent upper limit on the flux of fractional charge particles of $5\times10^{-14}$ particles/cm$^{2}$sec sr comes from the MACRO experiment \cite{macrofractional}.

\item {\bf Searches in bulk matter:  } The  third search method and the one that my colleagues and I are actively using \cite{mar,hal,dinesh} is searching for fractional charge particles in bulk matter. { \it This method depends upon the assumption that fractional charge particles were produced in the early universe \cite{perllee}and are present in the solar system.} One advantage of this method is that it covers a very large mass range, the upper limit being the same as that discussed in Sec. 3 for very massive particle searches. Another advantage of this method is that even a search with null results gives us an upper limit on the abundance of fractional charge particles in the solar system. Two different technologies are used, the levitometer method \cite{lev}and the liquid drop method. We use the latter method \cite{hen,perllee,dinesh}.
\end{itemize}

\subsubsection{ Summary of bulk matter searches for fractional charge particles}

The table below summarizes the major results of published searches for fractional charge particles in bulk matter. Note that 1 mg is about $6\times10^{20}$ nucleons.

\begin{center}
\begin{tabular}{|l|l|l|c|} \hline
\bf Method & \bf Experiment & \bf Material & \bf Sample Mass(mg)\\ \hline
\hline
superconducting levitometer & LaRue $et$ $al.$ \cite{lar} & niobium & 1.1\\ \hline
ferromagnetic levitometer & Marinelli $et$ $al.$ \cite{mari} & iron & 3.7\\ \hline
ferromagnetic levitometer & Smith $et$ $al.$ \cite{smi} & niobium & 4.9\\ \hline
ferromagnetic levitometer & Jones $et$ $al.$ \cite{jon} & meteorite & 2.8\\ \hline \hline
liquid drop & Joyce $et$ $al.$ \cite{joy} & sea water & .05\\
\hline
liquid drop & Savage $et$ $al.$ \cite{sav} & mercury & 2.0\\
\hline
liquid drop & Mar $et$ $al.$ \cite{mar} & silicone oil & 1.1\\
\hline
liquid drop & Halyo $et$ $al.$ \cite{hal} & silicone oil & 17.4\\
\hline
\hline
\end{tabular}
\end{center}

The only search that reported a positive result is that of LaRue {\it et al.} \cite{lar}, but Jones {\it et al.} \cite{jon} using a larger sample of niobium found no evidence for fractional charge particles.

In our recent search of 17.4 mg of silicone oil \cite{hal} we found no evidence for the existence of fractional charge particles. But among the $4.1\times10^{7}$ drops measured in this search there was one anomalous drop charge measurement. Ref. \cite{hal} gives full details. We plan to repeat the search in silicone oil with a larger sample.

\subsubsection{Our plans for future searches in bulk matter for fractional charge \\ particles}

As discussed by Lackner and Zweig \cite{lac} the presence of a fractional charge in an atom or molecule can drastically change the chemical properties of that atom or molecule. Therefore the most significant searches for fractional charge particles are in unrefined and unprocessed materials; this includes geochemical processing as well as human processing. Thus our plans for future searches for fractional charge particles in bulk matter are as follows:

\begin{enumerate}
\item {\bf Silicone Oil: }At present we are about to repeat our search in silicone oil with a larger sample. We recognize that this is a processed material, but feel that it is necessary to repeat our previous search \cite{hal}.
\item {\bf Terrestrial minerals: } Although earth materials have been subject to both melting and to geochemical processing, there are some ancient minerals of considerable interest \cite{lac}. We plan searches in such minerals.
\item {\bf Meteorites  from asteroids: }Asteroids consist of material that well represents the material in the solar system and yet has undergone a relatively small amount of natural processing. Meteorites from asteroids are easily obtained; we have acquired samples of the Allende meteorite for a search in the near future. {\it We believe that this will be our most significant search.}
\item {\bf Moon minerals:} There is some interest in searching for fractional charge particles in minerals form the moon's surface, but samples are scarce and we have no present plans for such a search.
\end{enumerate}

\subsubsection{Twenty-five year goals for searches for fractional charge \\  particles in bulk matter}

In accordance with the wishes of the organizers of this Workshop we give our twenty-five goals for our fractional charge research. Of course we hope that the plans just enumerated will lead to the discovery of fractional charge particles.

\begin{itemize}
\item {\bf Searches through larger samples:} At present we search through tens of milligrams of material and our present methods may reach gram size samples. We would like to develop methods that enable us to search through kilogram size samples. One way to do this would be to build hundreds of duplicate liquid drop machines \cite{hen}; we hope we can find a better way. 
\item {\bf Asteroids:} Our other long term goal is to obtain large samples directly from asteroids.
\end{itemize}

\section{SEARCHES FOR UNEXPECTED BEHAVIOR \\ OF KNOWN FORCES AND FOR NEW \\ FORCES}

\subsection{The strong, electromagnetic, and weak interactions}

Almost all experimental research on the strong, electromagnetic, and weak interactions has used the traditional methods of atomic, nuclear and particle physics. I include cosmic ray research in the traditional methods. However observational astrophysics and astronomy have contributed; for example the deficit in the sun's neutrino flux led to the investigation of neutrino oscillations.

I wonder what else we might learn from astrophysics and astronomy about these interactions? Certainly the traditional methods will continue to dominate, but we may come across puzzles in the phenomena of Fig. 1, puzzles that require a revision of our understanding of these interactions.

\subsection{The gravitational interaction}

This is the great adventure. At present we know a good deal about the classical aspects of the gravitational interaction \cite{gravity}:
\begin{itemize}
\item The equivalence principle holds to at lest $10^{-12}$ with respect to the material that comprises the earth.
\item The equivalence principle holds to at lest $10^{-3}$ with respect to the material that comprises our galaxy.
\item Observations of binary pulsars shows that gravitational wave radiation is explained by general relativity with a precision of at least $10^{-3}$.
\item Observations of binary pulsars have tested parts of strong field gravity to at least $10^{-3}$.
\end{itemize} 
 
Yet there is so much more that we want to learn about gravitation: how does it behave at very large distances, how does it behave at very small distances \cite{price}, are there modifications to general relativity? Figure 11 shows the  splendid and very ambitious experiments and observations that will go far toward answering these questions. But there are further questions: how can the quantum nature of gravity be observed, how can we do experiments on the quantum nature of gravity \cite{bra}? Here we must await new experimental ideas.

\begin{figure}[htb]
\begin{center}
\epsfig{file=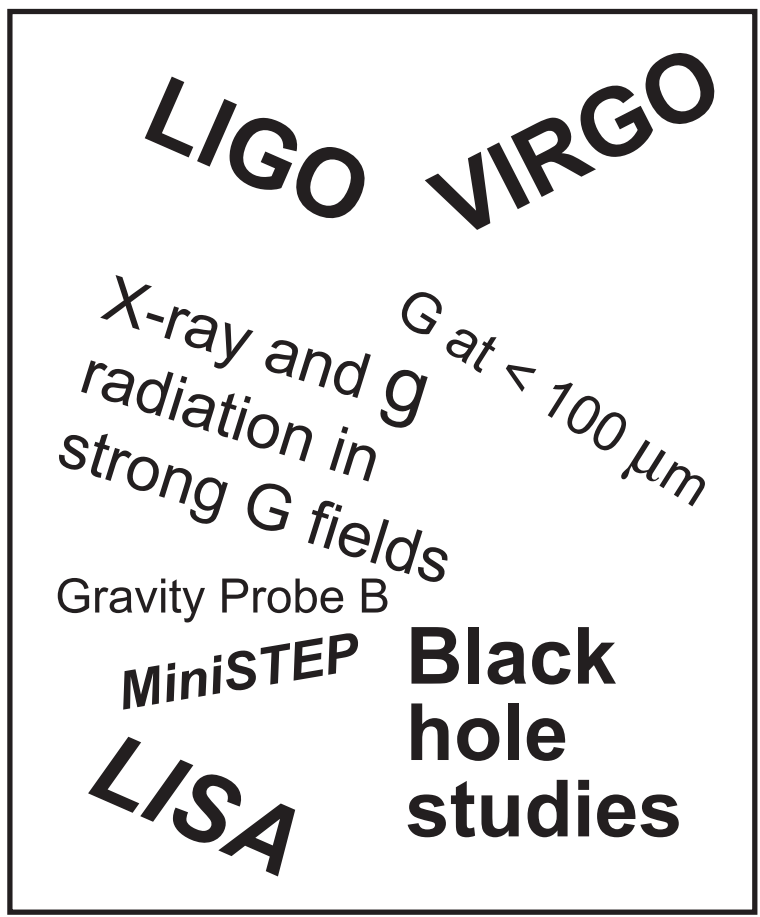,width=10cm}
\caption{The future of observational and experimental gravitational research.}
\end{center}
\end{figure}

\subsection{Are there undiscovered forces?}

This question haunts us. Are we so fortunate to have discovered all basic forces in the twentieth century? It seems to me that for physicists to say absolutely "yes" shows enormous conceit. Those who are less conceited will keep looking using the traditional methods of atomic and particle physics and the new methods of astrophysics and astronomy. 

\section{SEARCHES FOR NEW PHENOMENA AT VERY HIGH ENERGY}

The question here is whether there are astrophysical processes that contain reactions occurring at higher energies than are available, or will be available, at accelerators?  This question is prompted by the observation of very high cosmic rays, $10^{10}$ GeV and above. If these cosmic rays come from a reaction with a total energy above $10^{10}$ GeV, or if these cosmic rays come from the decay of particle with a mass above $10^{10}$ GeV/c$^{2}$, then there are astrophysical processes occurring at energies unreachable by current or planned accelerators. But if, as is more likely, these cosmic rays come from collective acceleration then there is no evidence for very high energy phenomena in astrophysical processes. Therefore in this final area, I am inclined to rely much more on the traditional accelerator research methods.

\section{ACKNOWLEDGEMENT}
I am very grateful to have had the opportunity to attend this Cosmic Genesis and Fundamental Physics Workshop. For someone such as I who has worked mostly in the traditional areas of particle physics, it opened wonderful vistas of new ways to investigate elementary particle physics.

This work was supported by the Department of Energy contract \\ DE-AC03-76SF00515.

\end{document}